\documentclass[fleqn,usenatbib]{mnras}

\usepackage{newtxtext,newtxmath}

\usepackage[T1]{fontenc}

\DeclareRobustCommand{\VAN}[3]{#2}
\let\VANthebibliography\thebibliography
\def\thebibliography{\DeclareRobustCommand{\VAN}[3]{##3}\VANthebibliography}

\usepackage{graphicx}	
\usepackage{amsmath}	
\usepackage{anyfontsize}
\usepackage[toc,page]{appendix}
\usepackage[hypcap=false]{caption}
\usepackage{ulem}
\usepackage{oplotsymbl}
\usepackage{MnSymbol}
\usepackage{xcolor}
\usepackage{subcaption}

\title[Nitrogenated Five-Membered Ring PAH Cations]{Infrared Spectroscopy of Nitrogenated Five-Membered Ring PAH Cations: Implications for the Aromatic Infrared Bands}

\author[Buragohain M.]{
Mridusmita Buragohain$^{1}$%
\thanks{Contact e-mail: \href{ms.mridusmita@gmail.com (MB),}
{ms.mridusmita@gmail.com}, ms.mridusmita@uohyd.ac.in}, 
Amit Pathak$^{2}$, \\ 
$^{1}$School of Physics, University of Hyderabad, Hyderabad 500 046, India\\
$^{2}$Department of Physics, Banaras Hindu University, Varanasi 221 005, India \\
}

\pubyear{\the\year{}}

\begin{document}
\label{firstpage}
\maketitle
\begin{abstract}
Despite the widely acknowledged role of polycyclic aromatic hydrocarbons (PAHs) and their variants as plausible carriers of the frequently observed mid-infrared emission bands across a wide range of astronomical targets, the contribution of five-membered ring PAHs and their nitrogenated derivatives, in particular, remains poorly investigated. Five-membered ring PAHs may arise as byproducts or intermediates during the formation and photochemical processing of larger PAHs. While the effects of nitrogenation on six-membered ring PAH cations have been studied previously, their five-membered ring counterparts remain relatively unexplored in the context of their potential contribution to the {Aromatic Infrared Bands (AIBs)}. In this work, we systematically investigate the infrared spectral characteristics of nitrogenated five-membered ring PAH cations in the 3--15~$\mu$m region, considering corannulene, circumcorannulene, and dicircumcorannulene as parent structures. To adequately account for the non-planar geometry and polarization effects in these molecules, their infrared spectra are calculated using DFT with the B3LYP functional and the cc-pVTZ basis set. We systematically examine the effects of N substitution on the principal infrared features by replacing C with N at different sites and assess the potential contribution of these species to the observed AIB spectrum.
\end{abstract}

\begin{keywords}
astrochemistry, ISM: molecules, (ISM:) dust, extinction, ISM: lines and bands, molecular data, software: simulations
\end{keywords}


\section{Introduction}
A variety of physical and chemical phenomena in the interstellar medium (ISM) validate the presence of interstellar polycyclic aromatic hydrocarbons (PAHs) {\citep[]{Tielens08, Li20},} which are carbonaceous gas-phase structures formed by fused aromatic benzene rings.
One such widely accepted phenomenon in favour of the existence of interstellar PAH is the `PAH-AIB' hypothesis, which suggests that interstellar PAH molecules absorb background UV photons and subsequently emit mid-infrared emission, popularly known as `Aromatic Infrared Bands (from hereonwards, AIBs)' through vibrational relaxation \citep[]{Leger84, Allamandola85, Allamandola89}. Within this framework, the C$-$H or C$-$C vibrational modes in PAHs are responsible for the prominent and frequently detected AIBs at 3.3, 6.2, 7.7, 8.6, 11.2, and 12.6~$\mu \rm m$.
These bands are characterised by variation in peak shifts, profiles, and relative intensities \citep[]{Hony01,Peeters02}.
Besides the conventional form of PAHs, variants of PAHs or structurally modified PAHs have been shown to successfully reproduce some of the weaker observed bands and source-to-source variations in the AIBs
identified in recent years using \textit{AKARI, Spitzer, ISO, SOFIA} and most recently, \textit{JWST} \citep[]{Peeters04, Hudgins05, Onaka14, Mridu15, Mridu16, Croiset16, Yang2020, Boersma2023, Draine25, Yang25}. 

In varying interstellar conditions, PAHs are expected to be chemically conditioned, leading to the {possible} formation of a diverse range of PAH variants and PAH-like species
(for example: hydrogenated PAHs \citep[]{Bernstein96}, nitrogenated PAHs \citep[]{Hudgins05},  deuterated PAH \citep[]{Peeters04}, cyano PAHs \citep[]{McGuire21}, metal-PAH complexes \citep[]{Simon07}, PAH clusters \citep[]{Rapacioli05} etc.).
Consequently, all forms of PAHs may collectively contribute to the observed AIBs, although some are expected to play a more significant role than others.
One notable example is a nitrogenated PAH (PANH), in which N replaces a C atom in the aromatic framework. PANH cations have been shown to reproduce the `Class A' 6.2~$\mu \rm m$ emission feature, with the closest agreement achieved when the nitrogen atom is incorporated within the interior of the PAH skeleton rather than at its periphery \citep[]{Hudgins05, Ricca2021}. {The observed variations in the peak positions and spectral profiles of AIBs across different astronomical environments have led to the classification of sources into Classes A, B, and C \citep[]{Peeters02}. In this classification, the 6.2~$\mu$m feature peaks near 6.22~$\mu$m for Class A sources, whereas Class B sources generally exhibit a redshifted peak between 6.24 and 6.28~$\mu$m. Class C sources exhibit a further redshift, with the corresponding feature appearing near 6.3~$\mu$m.} 

Recently discovered interstellar PAHs include various cyano-substituted compounds such as cyanonaphthalene, cyanoacenaphthylene, cyanopyrene, and cyanocoronene \citep[]{McGuire21,Cernicharo2024,Wenzel24,Wenzel25}. The cyano (CN) group imparts a permanent dipole moment to the otherwise nearly nonpolar PAH structure, facilitating their identification via rotational spectroscopy with radio telescopes. Conversely, the majority of unaltered PAHs exhibit significant symmetry and have minimal permanent dipole moments, making them predominantly undetectable by radio astronomical studies despite their probable prevalence in the interstellar medium. 

Another important class within the PAH family comprises five-membered ring PAHs, some of which may adopt nonplanar geometries possessing a permanent dipole moment. {These species are of particular astrophysical interest because such molecules are considered key intermediates in the top-down transformation of large PAHs into fullerenes \citep[]{Berne12}, such as C$_{60}$, C$_{70}$, both of which have been detected in the planetary nebula Tc 1 \citep[]{Cami10}. The ionized fullerene C$_{60}^+$ has also been identified in the ISM \citep[]{Campbell15}.} A bottom-up formation route from small PAHs to fullerenes has also been proposed by \citet[]{Gover21}, in which five-membered ring PAHs may occur as an intermediate structure. \citet{Burkhardt2021} reported the first detection of a five-membered ring PAH, indene, toward TMC-1 utilizing the Green Bank Telescope. Subsequently, its cyano-substituted derivative was also identified in the same source. The infrared spectra of five-membered ring PAHs have previously been investigated by \citet{Bausch15,Gauri20,Li24}. However, apart from a few, for example, unsubstituted or cyano (CN)-substituted species, the infrared spectroscopic properties of other derivatives of five-membered ring PAHs have received little attention. 
Nitrogenated six-membered ring PAHs (PANHs) have been extensively investigated owing to their ability to reproduce the `Class A' 6.2~$\mu$m emission feature. However, their five-membered ring counterparts have not been sufficiently investigated pertaining to the AIBs.

{\citet[]{Parker17} discussed the formation of nitrogenated PAHs (PANHs) through gas-phase reactions of nitrogen-bearing molecular precursors and subsequent aromatic ring growth, providing possible pathways for incorporating nitrogen directly into the carbon skeleton. Similarly, nitrogenated five-membered ring PAH cations may also form through molecular growth and energetic processing in the ISM. Depending on the physical and chemical conditions of the environment, five-membered ring PAHs can arise through gas-phase chemistry or structural rearrangement of PAHs \citep{Doddipatla21,Zhao21}, while nitrogen incorporation can produce PANH-like structures \citep{Hudgins05}. Subsequent UV-driven ionization may generate their cationic forms, although the detailed formation pathways remain poorly constrained. In contrast, cyano-substituted PAHs may form through the attachment of CN radicals to the edges of existing PAHs followed by hydrogen elimination without significantly altering their carbon skeleton \citep[]{Wenzel25a}. Although this provides a relatively straightforward formation pathway, PANHs may also form through nitrogen incorporation during aromatic ring growth rather than direct carbon replacement. Furthermore, the strong polarity of the CN group facilitates the detection of cyano-substituted PAHs through rotational spectroscopy. Although, nitrogenated six-membered or five-membered ring PAHs may possess permanent dipole moments, their rotational detectability depends on molecular symmetry, nitrogen-substitution position, and molecular size. Uncertainties in their formation pathways and interstellar abundances complicate their identification. Therefore, the detection of cyano-substituted PAHs, but not ring-nitrogenated PAHs, does not necessarily imply that peripheral CN attachment is intrinsically more favourable than nitrogen incorporation into the aromatic framework.}

This work presents the first systematic investigation of the infrared spectroscopic properties of nitrogenated five-membered ring PAH cations in the context of potential AIB carriers. Nitrogen incorporation into five-membered ring PAHs, particularly those with nonplanar geometries, is expected to enhance their permanent dipole moments, thereby improving their prospects for detection through rotational spectroscopy if they are present in the interstellar medium. \textit{JWST's} strong capability of detecting AIBs in diverse sources could be useful to further investigate the synthetic spectra.
\section{Computational Methodology and Objectives} 
Density functional theory (DFT) has become a standard computational tool for investigating the electronic and vibrational properties of molecules and has been extensively applied in the field of molecular astronomy to simulate the infrared spectra of interstellar PAHs and PAH-like species \citep{Mridu15,Bausch15, Mridu18, Mridu20,Li24,Mridu25}. 
In order to obtain an accurate spectrum to further compare with observations, a suitable combination of functional and basis set to be used on DFT calculation along with a scaling factor is required. The scaling factor accounts for the overestimation of theoretical frequencies in comparison to experimental frequencies. A previous study by \citet[]{Ricca2021} reported that b3lyp hybrid functional with cc-PVTZ basis set is more appropriate for the treatment of ploarized molecules, such as non-planer nitrogenated five-membered ring PAHs. For this study, DFT with the combination of b3lyp and cc-PVTZ basis sets has been adapted to generate the IR spectra of nitrogenated five-membered ring PAHs. Mode-dependent scaling factors of 0.964 for features below 5~$\mu \rm m$, 0.979 for the 5$-$10~$\mu \rm m$ region and 0.975 for features above 10~$\mu \rm m$ are applied to compensate for the overestimation of theoretical frequencies. Gaussian 16 \citep[]{g16} is used to perform the calculations.

An anharmonic calculation of sample PAHs is better suited to interpret the observed AIBs as such calculations result in overtones, combination bands and anharmonic shifts that are absent in the harmonic approximation. However, anharmonic calculations turn out to be computationally expensive if the sample molecules consist of more than a few C atoms. The smallest non-planar five-membered ring PAH considered in the present study contains 20 carbon atoms, making the anharmonic spectra computationally challenging. 
{Furthermore, substitution of a C atom by N at different sites generates several distinct structures for each parent PAHs considered here, substantially increasing the computational cost of performing anharmonic calculations for the complete molecular set.} In order to compensate for the anharmonic shifts and to convert theoretically computed absorption spectra into emission spectra, a 15 cm$^{-1}$ redshift is applied \citep[]{Bauschlicher08, Boersma13}  
{This report focuses more on band positions rather than intensity, and for wavelengths beyond $\sim$~6~$\mu$m, the band positions of scaled harmonic frequencies match well with both the anharmonic and NIST experimental data \citep[]{Mridu25}.} 
{Nevertheless, mode-dependent shifts in the vibrational frequencies introduced by anharmonicity may not be fully accounted for by a uniform empirical correction. Consequently, the precise positions of individual bands, including those associated with the five-membered ring skeletal vibrations, may differ from the present harmonic predictions.
Anharmonic effects are generally more pronounced} for features below 5~$\mu \rm m$ and for the overall study of the whole range, the adopted harmonic approach, together with the empirical scaling factor and redshift correction, is appropriate against the computationally demanding anharmonic calculations. The scaled, redshifted frequencies and the corresponding intensities are further converted into gaussian spectra with FWHM 30 cm$^{-1}$
for bands shortward of 9~$\mu \rm m$, 
10 cm$^{-1}$ for bands between 10 and 
15~$\mu \rm m$, and 5 cm$^{-1}$ for bands between 15 and 20~$\mu \rm m$. 
The FWHM is scaled linearly from 30$-$10~cm$^{-1}$ for
bands between 9 and 10~$\mu \rm m$. The selection of FWHM values are decided by the FWHM values of observed PAH spectra \citep[]{Peeters02, Ricca2021}.

\begin{figure*}
\centering
\includegraphics[width=13cm,height=4.5cm]{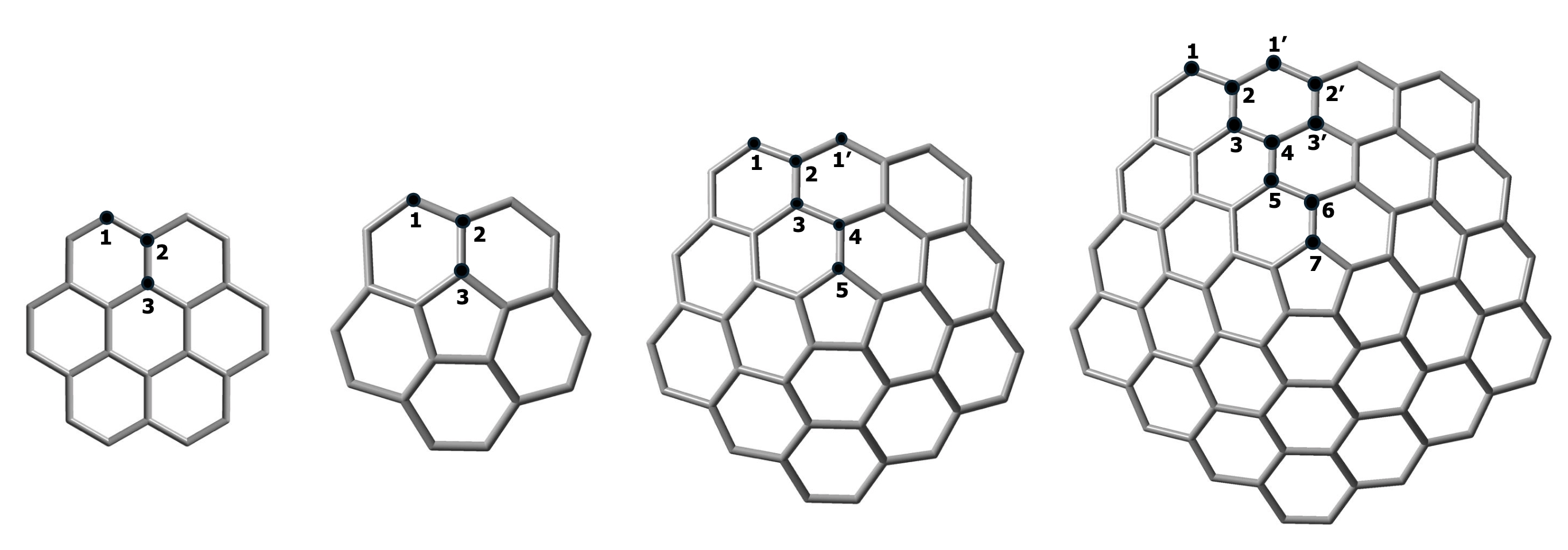}
\caption{{Coronene (C$_{24}$H$_{12}$),} Corannulene (C$_{20}$H$_{10}$), circumcorannulene (C$_{45}$H$_{15}$), and dicircumcorannulene (C$_{80}$H$_{20}$) 
(from left to right). The numbers inside the structures indicate the unique sites available for nitrogen substitution, with position 1 
corresponding to the outermost carbon atom and the highest-numbered position corresponding to the innermost carbon atom for nitrogen substitution. For N substitution at position 1, the attached H atom is removed and is referred to as exoskeletal-N
substitution, whereas remaining N substitution at inner positions is referred to as endoskeletal-N
substitution.} 
\label{fig1}
\end{figure*}

PAHs, selected for this study include cationic forms of corannulene (C$_{20}$H$_{10}$), circumcorannulene (C$_{45}$H$_{15}$) and circumcircumcorannulene (C$_{80}$H$_{20}$, from hereonward dicircumcorannulene) along with their nitrogenated variants, where a single N atom replaces a C atom at different positions starting from the external side and moving towards the inner core (Figure~\ref{fig1}). The unique position for N substitution is labelled as 1, 2, 3, .... etc. We refer to nitrogen substitution at the peripheral (outermost) carbon site as exoskeletal-N substitution, whereas substitution at the inner carbon sites is referred to as endoskeletal-N substitution. This terminology is used to distinguish the different N-substitution positions within the molecular framework.
Corannulene, circumcorannulene and dicircumcorannulene are five-membered ring counterparts of coronene (C$_{24}$H$_{12}$), circumcoronene (C$_{54}$H$_{18}$) and circumcircumcoronene (C$_{96}$H$_{24}$) respectively. Previous studies \citep[]{Hudgins05, Ricca2021} have investigated the infrared spectra of  nitrogenated six-membered ring PAH (PANH) cations to understand the influence of nitrogen substitution on the position of the 6.2~$\mu$m emission feature, which is generally attributed to PAH cations. Motivated by this, the present work examines the effect of nitrogen incorporation on the infrared spectra of five-membered ring PAH cations. Accordingly, the investigation is limited to the cationic forms of the selected molecules. While the manuscript intends to focus primarily on  the influence of nitrogen substitution on the 6.2~$\mu$m emission feature, the effect on other prominent infrared features is also studied. As for nitrogenation, for the external site (1 or 1$^\prime$), the adjacent H atom is removed to avoid the formation of an N$-$H bond, which would otherwise introduce a strong stretching feature near 2.9~$\mu$m. Since no corresponding astronomical emission feature is observed in this wavelength region, such N$-$H containing structures are excluded from the present study.  

{\citet{Hudgins05} suggested an N/C ratio of $\sim$1/30 for the PANH cation population, implying that more than one N atom may be incorporated into large six-membered ring PAH cations. As the nitrogenation-induced shift toward 6.2~$\mu$m decreases with increasing molecular size for singly nitrogenated species, multiple nitrogen substitutions may therefore be required to reproduce the `Class A' 6.2~$\mu$m feature. A similar scenario may also apply to large five-membered ring PAH cations, for which multiple nitrogen substitutions could further influence the position of the C--C$\rm_{stretching}$ feature. However, multiple N substitutions introduce a substantially larger number of possible structural isomers. As the present work represents an initial investigation of nitrogenated five-membered ring PAH cations in the context of the AIBs, we restrict the current study to singly nitrogenated species and defer a systematic investigation of multiple nitrogen substitutions to future work.}

Based on the calculated zero-point-corrected electronic energies, five-membered ring PAH cations (for example: corannulene cation) are generally less 
stable than their six-membered ring counterparts (for example: coronene cation). However, their stability increases with molecular size, progressing from corannulene to dicircumcorannulene cations, although it remains lower than that of the corresponding six-membered ring PAH cations. For the five-membered ring PAH cations considered here, nitrogen substitution further enhances their stability relative to the respective parent cations, although they remain less stable than the corresponding six-membered ring counterparts. The stability is weakly dependent on the position of the nitrogen atom within the molecular framework.  

{The permanent dipole moments of the PAH cations are important to evaluate their potential detectability through rotational spectroscopy, as well as their infrared spectral properties. We have calculated the centre-of-mass-corrected dipole moments of coronene and corannulene cations, and their singly nitrogenated analogues at the B3LYP/cc-pVTZ level of theory. The highly symmetric coronene cation has an almost zero dipole moment whereas the nonplanar corannulene cation has a permanent dipole moment of 2.38 D.  Substitution of a C atom by N at different positions within the molecular framework modifies the permanent dipole moment, with the calculated values following the increasing order:  coronene cation ($\approx$ 0 D) < corannulene cation (2.38 D) < N substituted coronene cation at position 3 (2.62 D) < N substituted coronene cation at position 1 (2.73 D) < N substituted corannulene cation at position 3 (2.79 D) < N substituted coronene cation at position 2 (3.60 D) < N substituted corannulene cation at position 1 (3.61 D) < N substituted corannulene cation at position 2 (4.01 D). The unique sites for nitrogen substitution (1, 2, and 3) considered for coronene correspond to the equivalent substitution sites in corannulene, as illustrated in Figure~\ref{fig1}. These results indicate that the magnitude of the permanent dipole moment strongly depends on the position of the nitrogen substitution and the consequent redistribution of electronic charge in the molecular framework. {A similar dependence of the dipole moment on nitrogen substitution position may also be expected for larger PAH cations, such as circumcoronene, circumcorannulene, dicircumcoronene, and dicircumcorannulene, and their nitrogenated variants, although further calculations are required to establish the corresponding trends.}
The presence of a five-membered ring introduces molecular curvature and an intrinsic dipole moment, but this does not necessarily result in a larger dipole moment than that of a nitrogenated six-membered ring PAH cation. 
The relatively large dipole moments of these nitrogenated PAH cations suggest that they may have observable rotational transitions, but their actual observability in the ISM also depends on molecular abundance, rotational excitation conditions, and the distribution of dipole-moment components over the principal inertial axes.}

\section{Results and Discussion}
This section discusses the infrared spectra of the selected molecules across different wavelength regions. These regions are selected based on the positions of the prominent observed AIBs, as well as the spectral ranges in which the calculated features are most pronounced. For each species, the result and discussions are divided into four wavelength ranges: (i) 3.3~$\mu$m, (ii) 6.2~$\mu$m, (iii) 7--10~$\mu$m, and (iv) beyond 10~$\mu$m. In each plot, the peak positions of the main AIB features at 3.3, 6.2, 7.7, 8.6, 11.2 and 12.7~$\mu$m are marked with steel blue dotted lines for reference. The intensities are made relative (int$\rm_{rel}$) to the peak intensity of the 6.2~$\mu \rm m$ feature.
\subsection{Corannulene}
\begin{figure*}
\includegraphics[width=8cm,height=9cm]{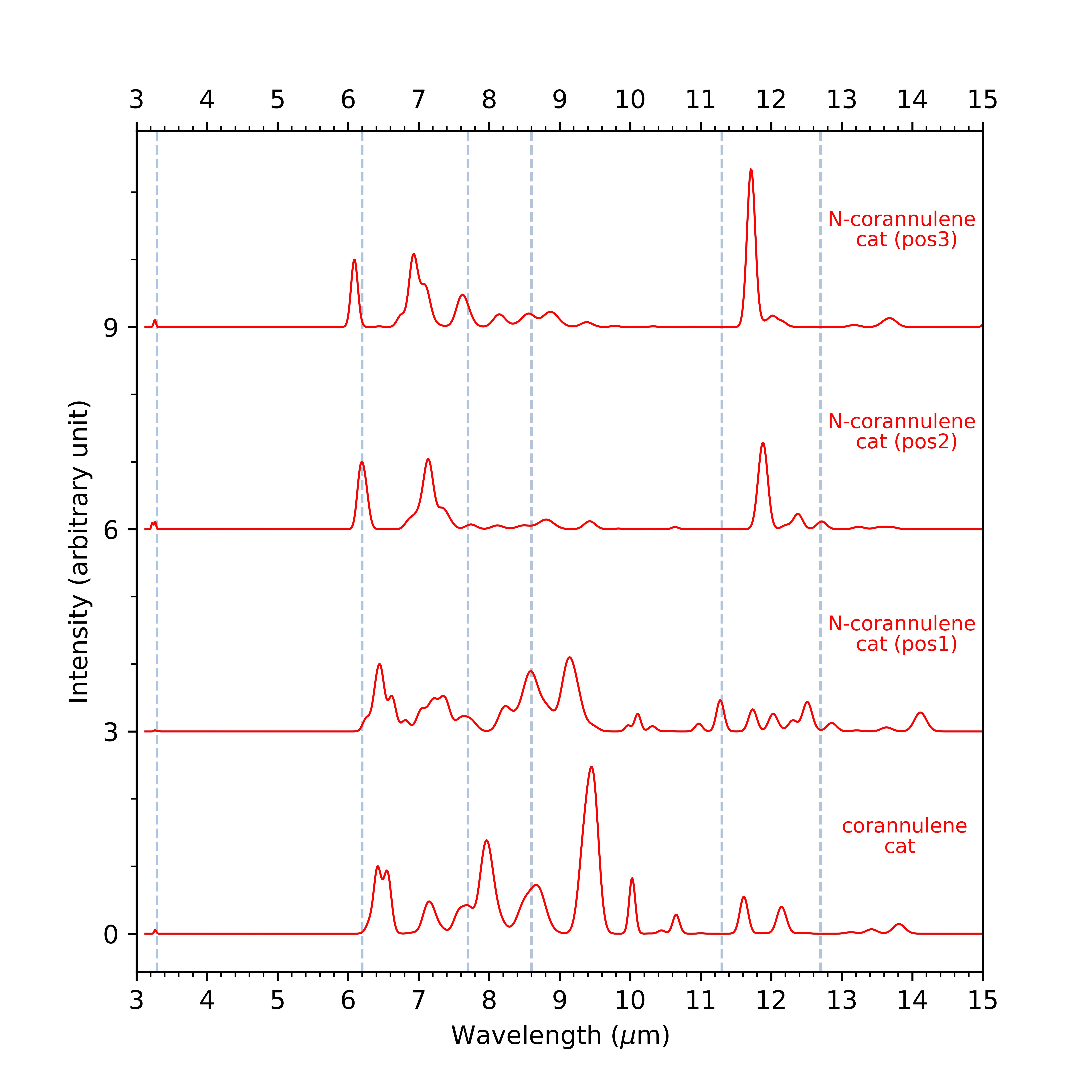}
\vspace{-1em}
\caption{IR spectra of corannulene cation and its nitrogenated variants. The cation is defined as `cat' in the label. The pos 1, pos 2 and pos 3 in the label denote the position where an N  atom is substituting a H atom.} 
\label{fig2}
\end{figure*}

Fig~\ref{fig2} shows the 3~$-$15~$\mu \rm m$ region IR spectra of corannulene cation (C$_{20}$H$_{10}^+$) and its nitrogenated variants. The cation and position of nitrogen in the PAH structure are defined as `cat' and `pos' in the figures. 
\subsubsection*{3.3~$\mu m$ region}
Six-membered ring PAH cations generally exhibit a weak C--H$\rm_{stretching}$ feature near 3.3~$\mu$m. The corannulene cation, a non-planar molecule containing a five-membered ring, is no exception. Nitrogen substitution in the corannulene cation does not affect the intensity of this feature, with only a marginal increase observed for N substitution at positions 2 and 3. The peak position remains essentially the same at 3.26~$\mu$m across all the variants of corannulene cation.
\subsubsection*{6.2~$\mu m$ region}
The 6.2~$\mu$m AIB is generally associated with aromatic C$-$C$\rm_{stretching}$ vibrations in PAHs, particularly cations. Observations suggest variation in the peak position of 6.2~$\mu$m AIB feature. Sources exhibiting this feature near 6.2~$\mu$m are classified as Class A, whereas those showing the feature near 6.4~$\mu$m are classified as Class C; sources with intermediate peak positions are classified as Class B.

For corannulene cation, there are two neighboring peaks at 6.42 ~$\mu$m (int$\rm_{rel}$1) and 6.55~$\mu$m (int$\rm_{rel}$~0.94), arising mainly from C--C$\rm_{stretching}$ and C--C--C skeletal vibration. For N substitution at position 1, the main feature appears at 6.45~$\mu$m with a blue shoulder at 6.31 ~$\mu$m ((int$\rm_{rel}$ 0.27) and another less intense feature peaking at 6.62~$\mu$m (int$\rm_{rel}$ 0.53). As N substitution moves towards the inner side of the structure, a blueshifted symmetric feature appears at 6.2 and 6.09~$\mu$m  for positions 2 and 3, respectively. A similar trend has been reported for nitrogenated six-membered ring PAH cations, where the 6.2~$\mu$m feature progressively shifts toward shorter wavelengths as the N atom moves from the outermost toward the innermost position. In the present scenerio of five-membered ring PAH cations, the dominant feature shifts from the Class C region toward the Class A position with increasing depth of N incorporation.
\subsubsection*{7$-$10~$\mu m$ region}
Corannulene cation 
exhibits several infrared features in the 7$-$10~$\mu$m region arising from C--H$\rm_{in-plane}$ bending coupled with C--C$\rm_{stretching}$ and C--C--C$\rm_{skeletal}$ deformation. 
Prominent features occur at approximately 7.15, 7.96, 8.67, 9.45, and 10.03~$\mu$m, with the 7.96~$\mu$m feature accompanied by a blue broad shoulder at 7.68~$\mu$m. The feature at 9.45~$\mu$m is the strongest in this region, followed by the feature at 7.96~$\mu$m.

The exoskeletal-nitrogenated corannulene (N at position 1) shows a broadly similar spectral pattern as that of corannulene cation, although with reduced intensities. Its prominent features are located at approximately 7.35, 7.65, 8.24, 8.59, and 9.14~$\mu$m, with the 9.14~$\mu$m feature being the strongest in this wavelength range. In contrast, N substitution at the inner positions (positions 2 and 3) substantially suppresses the spectral intensity between 7 and 9~$\mu$m. For position 2, an asymmetric feature with an intensity comparable to that of the 6.2~$\mu$m feature appears at 7.14~$\mu$m, while for position 3, a similar feature is shifted to 6.93~$\mu$m. In addition, N substitution at position 3 produces a moderately intense feature at 7.62~$\mu$m. These features are due to C--H$\rm_{in-plane}$ bending coupled with C--C$\rm_{stretching}$ and C--C--C$\rm_{skeletal}$ deformation. {Some of these vibrational modes also involve contributions from C–N stretching, which are strongly coupled with other skeletal motions, making it difficult to identify distinct C--N stretching features.}
\subsubsection*{Beyond 10~$\mu m$ region}
The corannulene cation exhibits four moderate-to-weak features at 10.03, 10.65, 11.61, and 12.15~$\mu$m. The first two features arise from C--H$\rm_{in-plane}$ bending modes coupled with C--C--C$\rm_{skeletal}$ deformations, whereas the latter two are attributed to C--H$\rm_{out-of-plane}$ or C--H$\rm_{oop}$ vibrations. For exoskeletal N substitution at position 1, several weak features (int$\rm_{rel}$ < 0.5) appear, with peaks at 10.11, 11.27, 11.74, 12.03, 12.52, and 14.12~$\mu$m. In contrast, N substitution at position 2 results in a prominent and sharp feature at 11.88~$\mu$m (int$\rm_{rel}$ $\sim$ 1.3), accompanied by two weaker features arising from C--H$\rm_{oop}$ bending. For N substitution at position 3, this prominent feature shifts toward a shorter wavelength of 11.71~$\mu$m and exhibits a substantial increase in intensity (int$\rm_{rel}$ $\sim$ 2.3). However, features with such high relative intensities at these positions are not typically observed in astronomical spectra.
\subsection{Circumcorannulene}
\begin{figure*}
\includegraphics[width=8cm,height=13cm]{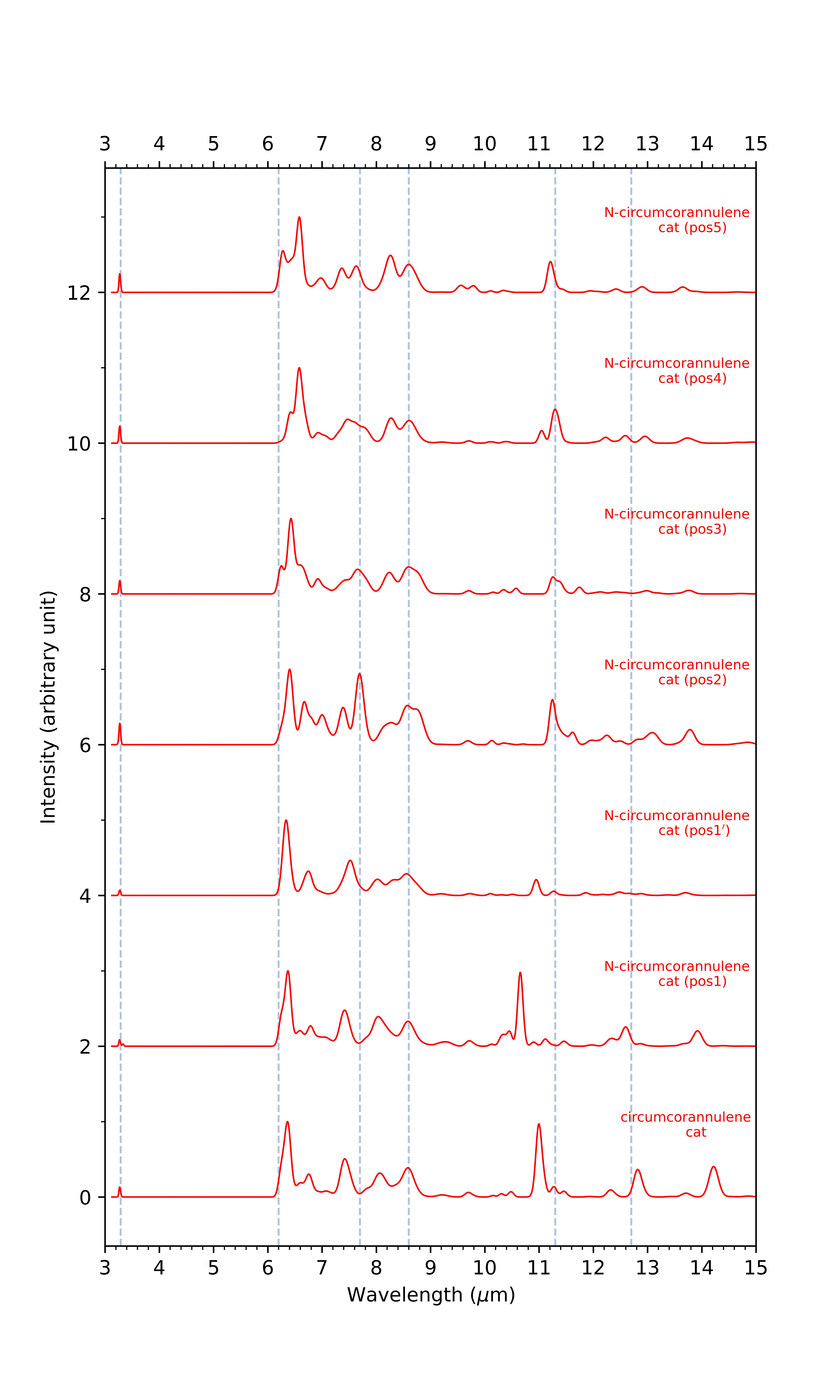}
\vspace{-3em}
\caption{IR spectra of circumcorannulene cation and its nitrogenated variants.} 
\label{fig3}
\end{figure*}

Fig~\ref{fig3} presents the 3~$-$15~$\mu \rm m$ IR spectra of circumcorannulene cation (C$_{45}$H$_{15}^+$) and its nitrogenated variants. N is placed at six different sites in the PAH structure, including two unique sites for substitution at the periphery (marked as 1 and 1$^\prime$ in figure~\ref{fig1}).
\subsubsection*{3.3~$\mu m$ region}
The 3.3~$\mu$m C--H$\rm_{stretching}$ feature is weakly present in the parent circumcorannulene cation (int$\rm_{rel}$ $\sim$ 0.13). N substitution at the peripheral positions (N at positions 1 and 1$^\prime$) produces similar intensities as that of the circumcorannulene cation. However, substitution at the inner positions of the circumcorannulene cation (positions 2--5) enhances the intensity of the C--H$\rm_{stretching}$ feature to int$\rm_{rel}\sim$~0.2-0.3. The peak position remains essentially invariant at the same wavelength of 3.27~$\mu$m for both the parent and nitrogenated variants.
\subsubsection*{6.2~$\mu m$ region}
For the circumcorannulene cation, the feature predominantly arising from C--C$\rm_{stretching}$ vibrations peaks at 6.37~$\mu$m (int$\rm_{rel}$ 1), close to the Class C `6.2~$\mu$m' position. 
The peak position and profile remain nearly unchanged for its nitrogenated counterparts with N at positions 1, 1$^\prime$ and 2. 
For N substitution at positions 3, 4, and 5, the main feature develops additional subpeaks, with the strongest peak appearing at 6.43 for position 3 and at 6.58~$\mu$m for positions 4 and 5. 
The corresponding subfeatures peak at 6.25, 6.42, and 6.28~$\mu$m for positions 3, 4, and 5, respectively.

For the circumcorannulene cation and its nitrogenated counterparts with N substitution at positions 1 and 1$^\prime$, a weaker feature (int$\rm_{rel}\sim$ 0.3) is present nearly at 6.8~$\mu$m arising from C--C$\rm_{stretching}$. For position 2, the feature (int$\rm_{rel}\sim$ 0.6) shifts to 6.7~$\mu$m, while for position 3, it appears as a red shoulder (int$\rm_{rel}\sim$ 0.4) at 6.6~$\mu$m to the main feature. For positions 4 and 5, this feature progressively blends with the main feature and is no longer distinctly resolved.
\subsubsection*{7$-$10~$\mu m$ region}
Several features are present in the 7--10~$\mu$m region for all the considered variants, most of which are less intense than the 6.2~$\mu$m feature. An exception is the feature near 7.7~$\mu$m for the nitrogenated variant with N at position 2, whose intensity is comparable to that of the 6.2~$\mu$m feature. These features mainly arise from C--H$\rm_{in-plane}$ bending coupled with C--C$\rm_{stretching}$ and C--C--C$\rm_{skeletal}$ {deformation, with additional contributions from C--N$\rm_{stretching}$ in some of the nitrogenated variants.}

The circumcorannulene cation and its exoskeletal nitrogenated variants with N at positions 1 and 1$^\prime$ exhibit broadly similar spectral characteristics in this region. A feature 
peaks at 7.42~$\mu$m for both the parent molecule and the position 1 variant and at 7.52~$\mu$m for position 1$^\prime$. Additional features are observed near 8.0 and 8.6~$\mu$m in all three variants.

With N substitution at positions 2--5, additional spectral features, although small in intensity emerge. A comparatively weak feature appears near 7~$\mu$m in all these variants. 
Another feature occurs near 7.4~$\mu$m, peaking at 7.42~$\mu$m for position 2 and 7.36~$\mu$m for position 5. For positions 3 and 4, this feature appears as a blue shoulder to the stronger neighbouring features at 7.67 and 7.47~$\mu$m, respectively. For position 2, a feature arises at 7.69~$\mu$m that has an intensity comparable to that of the 6.2~$\mu$m feature. A corresponding feature is also present at 7.63~$\mu$m for position 5, although with reduced intensity. The features at 7.67 and 7.47~$\mu$m for positions 3 and 4, respectively, can be regarded as analogous to this component.

A feature near 8~$\mu$m exhibits a systematic shift with the position of N substitution. For the circumcorannulene cation and the position-1 variants, it occurs near 8~$\mu$m, whereas N substitution at positions 2--5 shifts this feature toward longer wavelengths, to approximately 8.3~$\mu$m. Another relatively prominent feature occurs near 8.6~$\mu$m in all the considered variants of the circumcorannulene cation. This band shows little variation in peak position 
with the location of N substitution.
\subsubsection*{Beyond 10~$\mu m$ region}

In the wavelength region beyond 10~$\mu$m, the circumcorannulene cation exhibits three prominent features, arising predominantly from C--H$\rm_{oop}$ bending coupled with C--C--C$\rm_{skeletal}$ vibrations, at 11.00~$\mu$m (int$\rm_{rel}$ $\sim$ 0.97), 12.82~$\mu$m (int$\rm_{rel}$ $\sim$ 0.37), and 14.22~$\mu$m (int$\rm_{rel}$ $\sim$ 0.40), along with several weaker features. Upon N substitution at position 1, these features are blueshifted to 10.66~$\mu$m (int$\rm_{rel}$ $\sim$ 0.98), 12.59~$\mu$m (int$\rm_{rel}$ $\sim$ 0.26), and 13.93~$\mu$m (int$\rm_{rel}$ $\sim$ 0.20), respectively, and are accompanied by several additional weak features. 
For position 1$^\prime$, the overall intensities are substantially reduced, with only a weak feature of int$\rm_{rel}$~0.21 appearing at 10.95~$\mu$m.

For N substitution at position 2, a prominent feature appears at 11.25~$\mu$m (int$\rm_{rel}$~0.60), accompanied by a shoulder at 11.61~$\mu$m (0.16). Additional weaker features occur at 12.25~$\mu$m (0.12), 13.09~$\mu$m (0.16), and 13.77~$\mu$m (0.20). At position 3, the overall intensities are further reduced, with relatively weak features at 11.26~$\mu$m (0.23) and 11.74~$\mu$m (0.09). For position 4, a feature occurs at 11.30~$\mu$m (0.45), accompanied by a neighbouring peak at 11.05~$\mu$m (0.17) and several weaker features at longer wavelengths. For position 5, a symmetric feature appears at 11.21~$\mu$m (0.41), followed by weaker features toward longer wavelengths.

\subsection{Circumcircumcorannulene or dicircumcorannulene}
\begin{figure*}
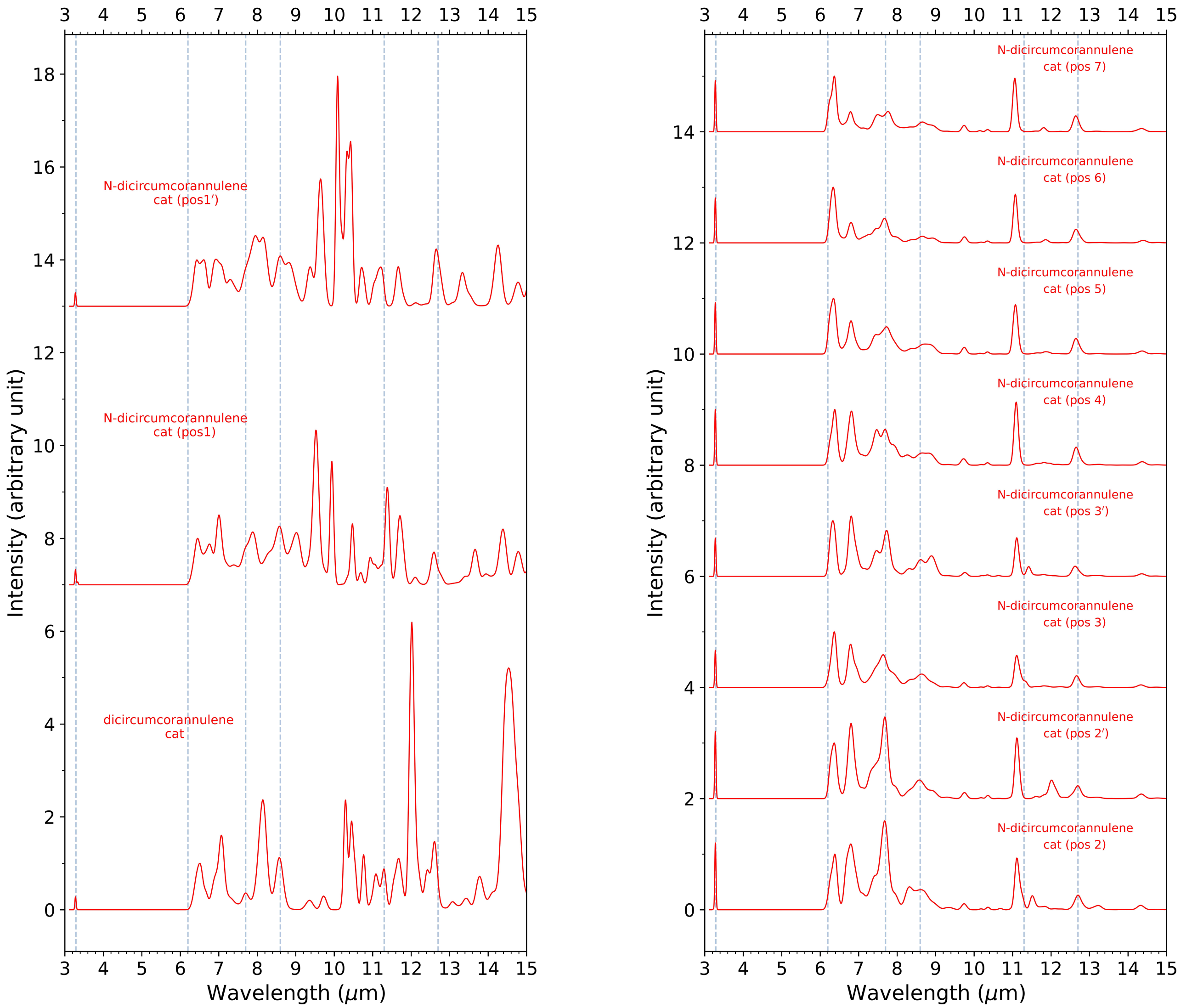

    \centering
    \begin{subfigure}{0.48\textwidth}
        \centering
        \includegraphics[width=\linewidth]{plot_C80H20_1.jpg}
        \label{fig4a}
    \end{subfigure}
    \hfill
    \begin{subfigure}{0.48\textwidth}
        \centering
        \includegraphics[width=\linewidth]{plot_C80H20_2.jpg}
        \label{fig4b}
    \end{subfigure}
    \vspace{-1cm}
    \caption{IR spectra of dicircumcorannulene cation and its nitrogenated variants. Left figure shows the spectra of the parent molecule and the first two variants (N at positions 1 and 1$^\prime$), while the right figure shows the spectra of the remaining variants (N at positions 2, 2$^\prime$, 3, 3$^\prime$, 4, 5, 6 and 7).} 
    \label{fig4}
\end{figure*}

Fig.~\ref{fig4} presents the 3--15~$\mu\rm m$ IR spectra of the dicircumcorannulene cation (C$_{80}$H$_{20}^{+}$) and its nitrogenated variants. N substitution is considered at ten distinct sites within the parent PAH structure including two unique peripheral carbon sites (1 and 1$^\prime$), as well as distinct carbon sites located at equivalent distances from the molecular periphery (2, 2$^\prime$, 3, and 3$^\prime$), as illustrated in Fig.~\ref{fig1}.

The dicircumcorannulene cation and its exoskeletal-nitrogenated variants with N at positions 1 and 1$^\prime$ exhibit several  unusually intense features, particularly at wavelengths longer than 10~$\mu$m. Such strong features, however, are not consistent with astronomical spectra. Similar spectral behavior was previously reported for the dicircumcorannulene cation by \citet{Bausch15}, who suggested employing a higher level of theory to further assess the IR spectra of such molecules.
However, our calculations using the higher-quality cc-pVTZ basis set, which is generally considered suitable for systems with significant polarization effects, yield similar spectral characteristics. To the best of our knowledge, this is the first report of the infrared spectral properties of nitrogenated dicircumcorannulene. Interestingly, the exoskeletal-nitrogenated variants exhibit spectral behavior similar to that of the parent dicircumcorannulene cation, including the occurrence of unusually intense features. Such extreme intensities are not observed for the remaining nitrogenated variants. 

Because of the unusually high intensities beyond 10~$\mu$m in the dicircumcorannulene cation and its nitrogenated variants with N substitution at positions 1 and 1$^\prime$, a detailed analysis of this wavelength region is beyond the scope of the present study. Henceforth, the analysis of these three variants is restricted to wavelengths shorter than 10~$\mu$m, whereas the full 3--15~$\mu$m range is considered for the remaining variants. The spectra are also divided into two sets across variants for ease of analysis and interpretation. 
\subsubsection*{3.3~$\mu m$ region}
For dicircumcorannulene cation and its nitrogenated variants with N substitution at positions 1 and 1$^\prime$, C--H$\rm_{ stretching}$ modes of int$\rm_{rel}\sim$~0.3 occurs consistently at 3.28~$\mu$m.  N incorporation at the inner positions from positions 2 to 7 significantly enhances the intensity of the 3.3~$\mu$m C--H$\rm_{ stretching}$ modes. In some cases (for example, N at positions 2 and 2$^\prime$), the intensities (int$\rm_{rel}\sim$1.2) even exceed the intensity of the 6.2~$\mu$m feature. 
The peak position and profile remain essentially unchanged at 3.28~$\mu$m across the different variants. 
\subsubsection*{6.2~$\mu m$ region}
For dicircumcorannulene cation, the spectral region beyond 6~$\mu$m contains an increasing number of relatively intense features. The parent dicircumcorannulene cation exhibits a prominent feature at 6.51~$\mu$m arising predominantly from C--C$\rm_{stretching}$ modes. For N substitution at position 1, the main feature appears at 6.45~$\mu$m, accompanied by another comparatively smaller feature at 6.77~$\mu$m, both arising from C--C$\rm_{stretching}$ modes. For position 1$^\prime$, this feature splits into two nearly equal-intensity subfeatures at 6.43 and 6.61~$\mu$m. In contrast, for N substitution at positions 2--7,
C--C$\rm_{stretching}$ modes occur as a symmetric profile at relatively constant positions of 6.38, 6.37, 6.37, 6.33, 6.38, 6.35, 6.34, and 6.37~$\mu$m, respectively. The position falls in the region of `class C' 6.2~$\mu$m feature. 
\subsubsection*{7--10~$\mu m$ region}
A prominent feature with int$\rm_{rel}$ 1.6 occurs at 7.07~$\mu$m in the dicircumcorannulene cation, arising from C--H$\rm_{in-plane}$ bending vibrations coupled with C--C$\rm_{stretching}$ modes. For N substitution at position 1, it occurs at 7~$\mu$m with int$\rm_{rel}$ 1.5.
For position 1$^\prime$, this feature shifts to 6.93~$\mu$m and has an intensity comparable to that of the 6.2~$\mu$m feature. For the remaining inner substitution sites, this feature shifts further toward shorter wavelengths, occurring near 6.8~$\mu$m, while its intensity progressively decreases toward the innermost positions. At position 2$^\prime$, its int$\rm_{rel}$ is 1.35, exceeding that of the 6.2~$\mu$m feature, whereas at the innermost position (position 7), it decreases to 0.36.

In the dicircumcorannulene cation, a relatively weak feature appears at 7.70~$\mu$m (int${\rm rel}$ $\sim$ 0.36), arising from C--H$\rm_{in-plane}$ bending vibrations coupled with C--C$\rm_{stretching}$ modes. This is followed by a considerably stronger feature at 8.14~$\mu$m (int${\rm rel}$ $\sim$ 2.36), attributed predominantly to C--H$\rm_{in-plane}$ bending modes. For N substitution at position 1, this mode appears at 7.89~$\mu$m (1.14). For position 1$^\prime$, this region contains two closely spaced features of nearly equal intensity at 7.96~$\mu$m (1.52) and 8.15~$\mu$m (1.48). With N substitution from position 2 toward the inner region, the feature initially gains some intensity and appears as an asymmetric profile at a position close to the position of the 7.7~$\mu$m AIB feature, attributed to C--H$\rm_{in-plane}$ bending vibrations coupled with C--C$\rm_{stretching}$ modes.
{Normal-mode analysis also reveals contributions from C--N stretching vibrations in several nitrogenated variants, coupled with C--C stretching and accompanied by C--C--C skeletal deformation, rather than appearing as isolated C--N stretching features, consistent with the behaviour observed for the previously discussed molecules.}
The main peak occurs at 7.68~$\mu$m for position 2 (int${\rm_{rel}}$ 1.60) and position 2$^\prime$ (1.47), at 
7.64~$\mu$m for position 3 (0.59) and 7.73~$\mu$m for position 3$^\prime$ (0.82), and at 7.69~$\mu$m for position 4 (0.64), at 7.73~$\mu$m (0.49) for position 5, at 7.67~$\mu$m (0.44) for position 6 and at 7.76~$\mu$m (0.37) for position 7.
These features are accompanied by neighbouring peaks or blue-shifted shoulder components near 7.4--7.5~$\mu$m , resulting in increasingly asymmetric spectral profiles.


A feature arising predominantly from C--H$\rm_{in-plane}$ bending modes, with an intensity comparable to that of the 6.2~$\mu$m feature, appears near 8.6~$\mu$m in the dicircumcorannulene cation and its exoskeletal-N substituted variants at positions 1 and 1$^\prime$. For N substitution at position 1$^\prime$, this feature is accompanied by another feature of the same vibrational origin at 8.82~$\mu$m (int$_{\rm rel}$ $\sim$ 0.94). In contrast, for N substitution at positions 2--7, the intensity of these modes decreases significantly, giving rise to a broader spectral profile characterized by multiple weaker subpeaks.

In addition, the dicircumcorannulene cation exhibits two weak features at 9.36~$\mu$m (int$\rm_{rel}$ 0.20) and 9.73~$\mu$m (0.29). These features arise from C--H$\rm_{in-plane}$ bending coupled with C--C--C$\rm_{skeletal}$ vibrations with some contribution from C--C$\rm_{stretching}$ modes. Upon N substitution at position 1, these modes become considerably stronger, giving rise to features at 9.02~$\mu$m (int$\rm_{rel}$ $\sim$ 1.12) and 9.52~$\mu$m (int$\rm_{rel}$ $\sim$ 3.33). For N substitution at position 1$^\prime$, the corresponding features occur at 9.37~$\mu$m (int$\rm_{rel}$ $\sim$ 0.85) and 9.64~$\mu$m (int$\rm_{rel}$ $\sim$ 2.74), respectively. Features of such high intensity at comparable wavelengths are not evident in astronomical spectra. In contrast, the remaining nitrogenated variants do not exhibit similarly strong features in this region. Instead, the inner N-substituted variants (positions 2--7) show a weak feature near 9.75~$\mu$m, with a relative intensity of approximately 0.11 arising from C--H$\rm_{in-plane}$ bending coupled with C--C--C$\rm_{skeletal}$ vibrations.

\subsubsection*{Beyond 10~$\mu m$ region}
Since the spectral features beyond 10~$\mu$m for the dicircumcorannulene cation and its exoskeletal-nitrogenated variants are unusually intense, and their theoretical treatment remains uncertain \citep[]{Bausch15}, this wavelength region is not considered further for these variants. Instead, the discussion focuses on the remaining N-substituted variants, which exhibit more moderate and physically plausible features beyond 10~$\mu$m.


A prominent feature attributed to solo C--H$\rm_{oop}$ bending modes occurs at 11.11~$\mu$m for N substitution at position 2, with an int$\rm_{rel}$ of 0.93. A feature at nearly the same wavelength is present in all the remaining variants, with int$\rm_{rel}$ values of 1.09, 0.58, 0.69, 1.13, 0.89, 0.87, and 0.96 for positions 2$^\prime$, 3, 3$^\prime$, 4, 5, 6, and 7, respectively. Thus, the peak position remains nearly invariant at $\sim$11.1~$\mu$m, while its intensity varies moderately with the N-substitution site.

An additional feature of similar vibrational origin occurs at 11.52~$\mu$m (int$\rm_{rel}$ $\sim$ 0.25) for N substitution at position 2 and is redshifted to 12.02~$\mu$m (int$\rm_{rel}$ $\sim$ 0.33) for substitution at position 2$^\prime$. Another feature appears at 11.42~$\mu$m (int$\rm_{rel}$ $\sim$ 0.17) for position 3$^\prime$, while for position 3 it manifests as a red shoulder to the prominent $\sim$11.1~$\mu$m feature. This feature progressively weakens for the remaining N-substituted variants.

A consistent feature near 12.7~$\mu$m is present across all the considered nitrogenated variants arising predominatly from duo C--H$\rm_{oop}$ bending modes. It peaks at 12.71~$\mu$m (int$\rm_{rel}$ 0.25) for position 2, 12.69~$\mu$m (0.23) for position 2$^\prime$, 12.66~$\mu$m (0.21) for position 3, 12.62~$\mu$m (0.18) for position 3$^\prime$, 12.66~$\mu$m (0.32) for position 4, 12.64~$\mu$m (0.28) for position 5, 12.64~$\mu$m (0.24) for position 6, and 12.64~$\mu$m (0.29) for position 7. Apart from this feature, only a few weaker bands are present, with no clear systematic effect attributable to the position of N substitution.

\section{Astrophysical Consequences and conclusion}
The primary objective of this study is to investigate the extent to which five-membered ring singly-nitrogenated PAH cations exhibit spectral characteristics analogous to those of six-membered ring nitrogenated PAH cations, particularly their ability to reproduce the Class A `6.2~$\mu$m' emission feature caused by C--C$\rm_{stretching}$. The effects of nitrogen incorporation on other infrared features across the 3--20~$\mu$m wavelength range are also examined.

Earlier, \citet[]{Hudgins1999, Hudgins05} suggested that with the growing size of six-membered ring PAH cations, the C--C$\rm_{stretching}$ features are blueshifted and even further blueshifted for its endoskeletal-nitrogenated counterparts. However, this trend of blueshifting of C--C$\rm_{stretching}$ feature with increasing molecular size appears to saturate for PAH cations containing more than $\sim$30 carbon atoms, with the dominant bands of larger PAHs clustering around $\sim$6.3~$\mu$m, corresponding to the Class C `6.2~$\mu$m' region. Also, the magnitude of this blueshift caused by single nitrogenation at different positions in the PAH structure decreases with increasing molecular size. In the present study, the molecular size is not varied through a simple linear sequence; instead, the selected species are obtained by progressively extending the corannulene framework in all directions. Furthermore, only single N substitution is considered for each molecular structure. For the molecules considered here, no systematic blueshift of the C--C$\rm_{stretching}$ feature is observed with increasing molecular size from corannulene to dicircumcorannulene. 
The feature is slightly blueshifted to 6.37~$\mu$m in circumcorannulene, whereas the dicircumcorannulene cation exhibits a redshift, with the dominant feature appearing at $\sim$~6.5~$\mu$m compared with $\sim$~6.4~$\mu$m for corannulene cation. 

As for nitrogenation, while the smaller five-membered ring PAH, corannulene, exhibits a systematic shift of the main feature toward the Class A `6.2~$\mu$m' region as the N atom is moved toward the molecular core, no such systematic trend is observed for the circumcorannulene cation. For N substitution at the peripheral positions (1 and 1$^\prime$) and position 2 of circumcorannulene, the spectra remain broadly similar to that of the parent molecule. As the N atom is incorporated toward the molecular core (positions 3--5), the main feature develops an increasingly asymmetric profile, with the dominant peak shifting further toward longer wavelengths. However, additional shorter-wavelength components emerge for substitution at these positions. For positions 3 and 5, these shorter-wavelength components occur closer to the Class B and Class C `6.2~$\mu$m' regions, respectively. Thus, although inward N substitution substantially modifies the spectral profile, it does not produce the same systematic blueshift toward the Class A position observed for corannulene. For the larger five-membered ring PAH, dicircumcorannulene cation, the primary C--C$\rm_{stretching}$ feature is slightly blueshifted from 6.51~$\mu$m to 6.45 and 6.43~$\mu$m upon N substitution at positions 1 and 1$^\prime$, respectively, along with additional components of comparable intensity toward longer wavelengths. 
As the N substitution moves inward from the molecular periphery toward the molecular core (positions 2--7), the primary C--C$\rm_{stretching}$ feature undergoes a further blueshift, converging to a nearly invariant position close to the Class C 6.2~$\mu$m region. 
However, the incremental blueshift associated with moving the N atom between successive inner substitution sites becomes progressively smaller toward the molecular core, indicating a reduced sensitivity of the C--C$\rm_{stretching}$ frequency to the precise N-substitution site in the inner region. This behaviour is consistent with the nitrogenation-induced spectral trends reported by \citet[]{Hudgins05}.

To examine the effect of nitrogenation on other prominent infrared features, we first consider the C--H$\rm_{stretching}$ mode near 3.3~$\mu$m. The feature remains largely unaffected in both position and relative intensity for corannulene cation and its nitrogenated counterparts. With increasing molecular size to circumcorannulene, N substitution at the inner positions (2--5) results in a small enhancement of the 3.3~$\mu$m feature, while its peak position remains nearly unchanged. 
For the larger dicircumcorannulene cation, the 3.3~$\mu$m feature is comparatively stronger than those of the smaller PAHs considered in this study, although its relative intensity remains lower than that of the 6.2~$\mu$m feature.
N substitution at the peripheral sites of dicircumcorannulene cation (positions 1 and 1$^\prime$) does not significantly alter the intensity of the 3.3~$\mu$m feature relative to the parent structure. In contrast, N substitution at the inner positions (2--7) leads to a substantial enhancement, with its intensity even exceeding that of the 6.2~$\mu$m feature for positions 2 and 2$^\prime$. Such a strong 3.3~$\mu$m feature is not supported by astronomical observations, thereby placing a constraint on the viability of these species as AIB carriers.

In the 7--10~$\mu$m complex region, corannulene cation and its exoskeletal-N substituted variant exhibit several features, whereas the number of features decreases for the remaining endoskeletal-N substituted variants. Although broad bands near 7.7 and 8.7~$\mu$m in corannulene cation coincide with prominent AIBs, stronger features near 8 and 9.5~$\mu$m lack clear astronomical counterparts. 
Moreover, for N substitution at positions 1, 2, and 3, strong features appear near 9.1, 7.14, and 6.9~$\mu$m, respectively, with intensities comparable to that of the 6.2~$\mu$m feature. Features of such high relative intensity at these wavelengths are not generally evident in astronomical spectra.

The features in the 7--10~$\mu$m region in circumcorannulene cation and its nitrogenated variants are mostly weaker than the 6.2~$\mu$m band. The 7.7~$\mu$m complex is blueshifted in the parent cation and peripheral N-substituted variants. For N substitution at position 2, a symmetric feature of comparable intensity to the 6.2~$\mu$m band appears near 7.7~$\mu$m, while for positions 3--5 it slightly blueshifts and develops a broader profile for positions 3 and 4. A weaker feature in comparison to
6.2~$\mu$m is present near 8.6~$\mu$m in all variants, coinciding with a prominent AIB position. However, an additional feature of comparable intensity as that of 8.6~$\mu$m is present between 8--8.3~$\mu$m in the respective molecules, which are not observationally viable.


For dicircumcorannulene cation and its exoskeletal-N substituted variants, stronger features than those near 7.7 and 8.6~$\mu$m occur at wavelengths where no comparably intense AIBs are observed, limiting their contribution as potential AIB carriers. The spectra become less complex for variants with N substitution at positions 2--7. 
However, the presence of a prominent feature near 7~$\mu$m (int$\rm_{rel} \gtrsim $ 1) 
still constrains their viability as AIB carriers.  The intensity of this feature is reduced for the inner positions (positions 5--7) and it is comparable to that of the 7.7~$\mu$m complex. This feature with int$\rm_{rel} \gtrsim $ 1 is also present for dicircumcorannulene cation  and its exoskeletal-N substituted variants.
A feature near 7.7~$\mu$m is present in the endoskeletal-N substituted variants; however, its intensity decreases and its profile becomes increasingly asymmetric as the N-substitution site moves toward the molecular core.  An 8.6~$\mu$m feature with int$\rm_{rel} \sim $ 1 is present in dicircumcorannulene cation and its exoskeletal-N substituted variants (N at positions 1 and 1$^\prime$), although for endoskeletal-N substituted variants (N at positions 2--7), the feature appears broad with much smaller intensity than that of 6.2~$\mu$m feature.

The 11.2~$\mu$m feature is generally associated with the C--H$\rm_{oop}$ bending mode of solo C--H groups in neutral PAHs \citep[]{Mridu25}, where a solo C--H group refers to a C--H bond without any neighboring C--H bond on the same aromatic ring. Corannulene does not contain solo C--H groups and, therefore, a feature associated with solo C--H$\rm_{oop}$ bending is not expected. Nevertheless, the corannulene cation exhibits a symmetric, moderate-intensity C--H$\rm_{oop}$ feature arising from duo C--H groups, i.e., C--H groups with one neighboring C--H bond on the same aromatic ring. This feature occurs near the wavelength typically associated with the C--H$\rm_{oop}$ bending modes of duo C--H groups in neutral PAHs. The exoskeletal-N substituted corannulene cation (N at position 1) possesses a solo C--H group, whose out-of-plane bending mode manifests as a moderate feature at 11.27~$\mu$m, coupled with C--C--C$\rm_{skeletal}$ motions. Another feature at 11.74~$\mu$m arises from the C--H$\rm_{oop}$ bending modes of the duo C--H groups. Although PAH cations generally exhibit C--H$\rm_{oop}$ features weaker than the 6.2~$\mu$m band, the endoskeletal-N substituted corannulene cations with N at positions 2 and 3 exhibit prominent duo C--H$\rm_{oop}$ features at approximately 11.9 and 11.7~$\mu$m, respectively, with intensities comparable to or exceeding that of the 6.2~$\mu$m feature.

The circumcorannulene cation contains both solo and duo C--H groups, whose out-of-plane bending modes give rise to features near 11 and 12.8~$\mu$m, respectively. Notably, the 11~$\mu$m feature has an intensity comparable to that of the 6.2~$\mu$m feature, which is generally not expected for PAH cations. 
For the N substitution at position 1, an intense feature is present nearly at 10.7~$\mu$m due to solo C--H$\rm_{oop}$ and C--C--C$\rm_{skeletal}$ vibrational modes. The strong feature due to solo C--H$\rm_{oop}$ in these two variants lies slightly away from the observed position of 11.2~$\mu$m AIB feature.
The vibrational modes at this wavelength range are substantially weaker for the N substitution at position 1$^\prime$. For the remaining endoskeletal-N substituted variants, the solo C--H${\rm oop}$ modes occur near 11.3~$\mu$m with intensities lower than that of the 6.2~$\mu$m feature, while the duo C--H${\rm oop}$ modes are even weaker than the solo C--H${\rm oop}$ modes.


The dicircumcorannulene cation and its exoskeletal-N substituted variants exhibit unusually intense features beyond 10~$\mu$m, in contrast to the other PAH species considered in this study. For the remaining N-substituted variants (positions 2--7), the spectra are comparatively less complex. While the dicircumcorannulene cation and its exoskeletal-nitrogenated variants possess open-shell electronic structures, the remaining endoskeletal-nitrogenated variants are closed-shell in nature. Consistent with the previous findings of \citet[]{Bausch15}, these results suggest that the chosen level of theory and basis set may have limitations in accurately describing the vibrational properties of very large, non-planar, open-shell PAHs containing five-membered rings.

Endoskeletal-N substituted variants (positions 2--7) show consistent features near 11.1 and 12.7~$\mu$m attributable to the out-of-plane bending modes of solo and duo C--H groups, respectively. While the 12.7~$\mu$m feature is relatively weak, the 11.1~$\mu$m feature has an intensity comparable to that of the 6.2~$\mu$m feature, which is not typically observed for conventional PAH cations. 

To summarize the main findings of this study, the key results are as follows:
\begin{enumerate}
    \item In five-membered ring PAH cations, the C--C$\rm_{stretching}$ feature does not exhibit a systematic size-dependent blueshift from corannulene to dicircumcorannulene; instead, a redshift is observed for dicircumcorannulene. 
    \item Nitrogenation produces a clear position-dependent blueshift in corannulene as the N-substitution site moves inward toward the molecular core. However, this trend becomes progressively weaker with increasing molecular size and does not shift the C--C$\rm_{stretching}$ feature of the larger species to the position of the Class A `6.2~$\mu$m' AIB. In the dicircumcorannulene cation, the position of the C--C$\rm_{stretching}$ feature becomes nearly invariant for N substitution at sites located toward the molecular core.
    \item N incorporation in a larger five-membered ring PAH cation, say dicircumcorannulene cation, substantially enhances the 3.3~$\mu$m C--H$\rm_{stretching}$ feature, in some cases exceeding the 6.2~$\mu$m intensity, which is inconsistent with observations. 
    \item In the 7--10~$\mu$m region, several additional bands are present besides the prominent 7.7 and 8.6~$\mu$m features, some of which exhibit intensities even greater than those of the established AIB features.
\item Nitrogenation of five-membered ring PAH cations significantly modifies the C--H$\rm_{oop}$ region. Endoskeletal-N substituted corannulene cations exhibit strong duo C--H$\rm_{oop}$ features at 11.7--11.9~$\mu$m, with intensities comparable to or exceeding that of the 6.2~$\mu$m feature. In endoskeletal-N substituted circumcorannulene cations, the solo and duo C--H$\rm_{oop}$ features remain weaker than the 6.2~$\mu$m feature, whereas endoskeletal-N substituted variants of the dicircumcorannulene cation exhibit solo C--H$\rm_{oop}$ features with intensities comparable to that of the 6.2~$\mu$m feature.
\end{enumerate}


While some of the theoretically predicted features of five-membered ring PAH cations and their nitrogenated variants, discussed above, may account for individual interstellar PAH emission features, the presence of unusually strong bands at wavelengths without clear astronomical counterparts raises questions about their viability as AIB carriers. The high spectral resolution and sensitivity of JWST, together with future observations from the Nancy Grace Roman Space Telescope, may help assess the presence of these features in astronomical sources. {In addition, a systematic investigation of multiply N-substituted five-membered ring PAH cations is important for further understanding the effects of nitrogenation on their spectral characteristics and their potential contribution to the AIBs.}

\section{ACKNOWLEDGEMENTS}
{MB thanks the Department of Science and Technology (DST), India for awarding her ’Innovation in Science Pursuit for Inspired Research (INSPIRE) Faculty fellowship’ and research grant. AP acknowledges financial support from the Institute of Eminence (IoE) grant of Banaras Hindu University (R/Dev/D/IoE/Incentive/2021- 22/32439),  financial support from the Department of Science and Technology-Japan Society for the Promotion of Science (DST-JSPS) grant 2024. MB and AP thank the Inter-University Centre for Astronomy and Astrophysics (IUCAA), Pune for visiting associateship. MB also thanks the Centre for Modelling, Simulation and Design (CMSD), University of Hyderabad, for providing high-performance computing facilities. The authors acknowledge the use of ChatGPT (OpenAI, GPT-5.6) for language refinement and improvement of the manuscript’s clarity. All revisions were reviewed and approved by the authors.}
\section{DATA AVAILABILITY}
The data underlying this article are available from the authors upon request.

\def\aj{AJ}%
\def\actaa{Acta Astron.}%
\def\araa{ARA\&A}%
\def\apj{ApJ}%
\def\apjl{ApJ}%
\def\apjs{ApJS}%
\def\ao{Appl.~Opt.}%
\def\apss{Ap\&SS}%
\def\aap{A\&A}%
\def\aapr{A\&A~Rev.}%
\def\aaps{A\&AS}%
\def\azh{AZh}%
\def\baas{BAAS}%
\def\bac{Bull. astr. Inst. Czechosl.}%
\def\caa{Chinese Astron. Astrophys.}%
\def\cjaa{Chinese J. Astron. Astrophys.}%
\def\icarus{Icarus}%
\def\jcap{J. Cosmology Astropart. Phys.}%
\def\jrasc{JRASC}%
\def\mnras{MNRAS}%
\def\memras{MmRAS}%
\def\na{New A}%
\def\nar{New A Rev.}%
\def\pasa{PASA}%
\def\pra{Phys.~Rev.~A}%
\def\prb{Phys.~Rev.~B}%
\def\prc{Phys.~Rev.~C}%
\def\prd{Phys.~Rev.~D}%
\def\pre{Phys.~Rev.~E}%
\def\prl{Phys.~Rev.~Lett.}%
\def\pasp{PASP}%
\def\pasj{PASJ}%
\def\qjras{QJRAS}%
\def\rmxaa{Rev. Mexicana Astron. Astrofis.}%
\def\skytel{S\&T}%
\def\solphys{Sol.~Phys.}%
\def\sovast{Soviet~Ast.}%
\def\ssr{Space~Sci.~Rev.}%
\def\zap{ZAp}%
\def\nat{Nature}%
\def\iaucirc{IAU~Circ.}%
\def\aplett{Astrophys.~Lett.}%
\def\apspr{Astrophys.~Space~Phys.~Res.}%
\def\bain{Bull.~Astron.~Inst.~Netherlands}%
\def\fcp{Fund.~Cosmic~Phys.}%
\def\gca{Geochim.~Cosmochim.~Acta}%
\def\grl{Geophys.~Res.~Lett.}%
\def\jcp{J.~Chem.~Phys.}%
\def\jpca{J.~Phys.~Chem.~A}%
\def\jgr{J.~Geophys.~Res.}%
\def\jqsrt{J.~Quant.~Spec.~Radiat.~Transf.}%
\def\memsai{Mem.~Soc.~Astron.~Italiana}%
\def\nphysa{Nucl.~Phys.~A}%
\def\physrep{Phys.~Rep.}%
\def\physscr{Phys.~Scr}%
\def\planss{Planet.~Space~Sci.}%
\def\procspie{Proc.~SPIE}%
\let\astap=\aap
\let\apjlett=\apjl
\let\apjsupp=\apjs
\let\applopt=\ao



\bibliographystyle{mnras}
\bibliography{mridu} 
\end{document}